# Zeta Functions with Dirichlet and Neumann Boundary Conditions for Exterior Domains

*J.-P. Eckmann*[1,2] *and C.-A. Pillet*[1]

[1]Dépt. de Physique Théorique, Université de Genève, CH-1211 Genève 4, Switzerland
[2]Section de Mathématiques, Université de Genève, CH-1211 Genève 4, Switzerland

**Abstract.** We generalize earlier studies on the Laplacian for a bounded open domain $\Omega \in \mathbf{R}^2$ with connected complement and piecewise smooth boundary. We compare it with the quantum mechanical scattering operator for the exterior of this same domain. Using single layer and double layer potentials we can prove a number of new relations which hold when one chooses *independently* Dirichlet or Neumann boundary conditions for the interior and exterior problem. This relation is provided by a very simple set of $\zeta$-functions, which involve the single and double layer potentials. We also provide Krein spectral formulas for all the cases considered and give a numerical algorithm to compute the $\zeta$-function.

## 1. Introduction

In an earlier paper [EP2], we derived an identity between the integrated density of states for the eigenvalues for the Laplacian in a domain $\Omega$ with Dirichlet boundary conditions, and the scattering phases for the exterior of the same domain, also with Dirichlet boundary conditions. In the present paper, we derive similar identities, and prove several of them, for the case where the boundary conditions can also be of Neumann type. We will discuss and illustrate similarities and differences between the various cases.

Although it is not possible to really formulate the identities we are going to derive without making precise definitions, we summarize here the main results in an informal way. We consider a bounded domain $\Omega$ in $\mathbf{R}^2$, and we let $\Gamma$ be its boundary. We denote by $G_z$ the free Green's function $G_z = (-\Delta - z)^{-1}$. This is an analytic function of $z$, except for a logarithmic singularity at the origin. We assume that the branch cut is along $\mathbf{R}^+$. We let $G_z(x, x')$, with $x, x' \in \mathbf{R}^2$, denote the integral kernel of $G_z$. Then we define

$$A_z = G_z|_{\Gamma \times \Gamma}, \quad B_z = -G_z N \cdot \nabla|_{\Gamma \times \Gamma}, \quad C_z = -N \cdot \nabla G_z N \cdot \nabla|_{\Gamma \times \Gamma},$$

where the normal $N$ points out of $\Omega$ and $|_{\Gamma \times \Gamma}$ is the restriction to the boundary. This is a more precise notation for the "normal derivative on the boundary." Setting $E_0 = -1$, we define 4 $\zeta$-functions, for $z \notin \mathbf{R}^+$,

$$\begin{aligned}
\zeta_{\mathbf{DD}}(z) &= \det\left(A_{E_0}^{-1} A_z\right), \\
\zeta_{\mathbf{DN}}(z) &= \det\left((\tfrac{1}{2} - B_{E_0})^{-1}(\tfrac{1}{2} - B_z)\right), \\
\zeta_{\mathbf{ND}}(z) &= \det\left((\tfrac{1}{2} + B_{E_0})^{-1}(\tfrac{1}{2} + B_z)\right), \\
\zeta_{\mathbf{NN}}(z) &= \det\left(C_{E_0}^{-1} C_z\right).
\end{aligned}$$

The symbols $\mathbf{D}$ and $\mathbf{N}$ stand for Dirichlet and Neumann boundary conditions. Then we have the following nice identities between the integrated densities of states $N_{\mathbf{B}}$ for the $-\Delta$ in the interior and the scattering phases $\Theta_{\mathbf{B}'}$ for the exterior $\Omega^c$ of $\Omega$, [1] with boundary conditions $\mathbf{B}, \mathbf{B}' \in \{\mathbf{D}, \mathbf{N}\}$,

$$\pi N_{\mathbf{B}}(E) = \Theta_{\mathbf{B}'}(E) - \operatorname{Im} \log \zeta_{\mathbf{B},\mathbf{B}'}(E + i0). \tag{1.1}$$

Furthermore,

$$|\operatorname{Im} \log \zeta_{\mathbf{B},\mathbf{B}'}(E + i0)| \leq \operatorname{const.} \mathrm{E}^{1/2} \log \mathrm{E}. \tag{1.2}$$

We also have a Krein trace formula, valid for any nice function $F$:

$$\operatorname{Tr}\big(F(-\Delta_{\Omega,\mathbf{B}} \oplus -\Delta_{\Omega^c,\mathbf{B}'}) - F(-\Delta)\big)$$
$$= \sum_n F(\lambda_{n,\mathbf{B}}) + \frac{1}{\pi i} \int dE \, \frac{dF(E)}{dE} \Theta_{\mathbf{B}'}(E), \tag{1.3}$$

---

[1] We define the scattering phase by $\det S_{E,\mathbf{B}'} = \exp(-2i\Theta_{\mathbf{B}'}(E))$, where $S_{E,\mathbf{B}'}$ is the S-matrix with boundary condition $\mathbf{B}'$.

where the $\lambda_{j,\mathbf{B}}$ are the eigenvalues of $-\Delta_{\Omega,\mathbf{B}}$, the Laplacian in $\Omega$ with boundary conditions $\mathbf{B}$. The S-matrix $S_{E,\mathbf{B}'}$ with boundary condition $\mathbf{B}'$ can also be described purely in terms of $A_z$, $B_z$, or $C_z$. We also give identities for the interacting Green's functions in 2 cases:

$$\begin{aligned} G_{\mathbf{DD}} &= G - G\gamma^* A^{-1}\gamma G\,, \\ G_{\mathbf{NN}} &= G - G\gamma_N^* C^{-1}\gamma_N G\,. \end{aligned} \quad (1.4)$$

Here $\gamma$ is the restriction to $\Gamma$, and $\gamma_N$ is the normal derivative on $\Gamma$. Note that one of the consequences of Eq.(1.4) is a *pointwise* spectral duality, see Theorem 3.3 below. Finally, we describe an algorithm for computing the $\zeta$-function which is based on the double layer potential $B_z$.

This paper can be read in two ways. On the one hand, it can be viewed as a collection of identities relating Dirichlet, and Neumann scattering to Dirichlet and Neumann membranes in terms of single and double layer potentials on the boundary of $\Omega$, through certain $\zeta$-functions. Many of these identities, although reminiscent of work by [KR, HS, SU, BS2] are in fact new. On the other hand, we give detailed proofs for those identities involving the double layer potential, and relating Dirichlet conditions on one side of the boundary to Neumann conditions on the other.

**Acknowledgments.** Our interest in the problems discussed in this paper has been provoked by the papers of Smilansky et al. [SU] and Steiner et al. [BS2]. Part of our findings are a direct consequence of fruitful discussions with and inspiring seminars by members of these groups. In addition, we have profited from helpful discussions with V. Ivrii, A. Jensen, V.S. Buslaev and D.R. Yafaev. They all have contributed to clarify our views about the relevant issues. This work was supported by the Fonds National Suisse, and many of our contacts have been made possible by the semester "Chaos et quantification" at the Centre Emile Borel in Paris.

## 2. Notations and Definitions

We consider domains $\Omega$ which we call "standard domains."

**Definition.** A domain $\Omega \in \mathbf{R}^2$ is called a *standard domain* if
 a) $\Omega$ is bounded.
 b) $\Gamma = \partial\Omega$ is piecewise $\mathcal{C}^2$, with a finite number of pieces.
 c) The angles at the corners are non-degenerate, i.e., neither 0 nor $2\pi$.
 d) The complement $\Omega^c$ of $\Omega$ is connected.

**Remarks.**
 – It should be noted that the definition allows for domains which consist of several pieces. We shall, however only deal with the case of a connected domain to keep the notation simple.
 – The theory would be somewhat easier, with bounds which are not really any better, if we restricted our attention to smooth domains. However, in view of applications and examples, we think that the inclusion of corners is important.

To formulate our results, we need to define the various spectral densities and scattering phase shifts.

**Notation.** We shall use throughout the subscripts $\mathbf{D}$ and $\mathbf{N}$ to denote Dirichlet and Neumann boundary conditions, respectively. We denote by $\mathbf{B}, \mathbf{B}'$ a choice of boundary conditions among $\{\mathbf{D}, \mathbf{N}\}$.

**Definitions.** We define here the quantities $N_\mathbf{B}$, $\Theta_{\mathbf{B}'}$. Let $\mathbf{B}, \mathbf{B}'$ be boundary conditions in $\{\mathbf{D}, \mathbf{N}\}$.
- The quantity $N_\mathbf{B}(E)$ denotes the number of eigenvalues below $E$, counted with multiplicity, of $-\Delta_{\Omega, \mathbf{B}}$. Here, $\Delta_{\Omega, \mathbf{B}}$ is the Laplacian in $\Omega$ with boundary condition $\mathbf{B}$ on $\Gamma$.
- The quantity $\Theta_{\mathbf{B}'}(E)$ is the total scattering phase for the scattering operator in $\Omega^c$, with boundary conditions $\mathbf{B}'$ on $\Gamma$. I.e., $\det S_{E, \mathbf{B}'} = e^{-2i\Theta_{\mathbf{B}'}(E)}$. It is normalized to $\Theta_{\mathbf{B}'}(0) = 0$, and it is defined as a *continuous* function of $E$.

Our analysis will be based on a study of the single and double layer potentials which we define next. We denote by $G$ the Green's function

$$G_z = \frac{1}{-\Delta - z}, \quad z \in \mathbf{C} \setminus \mathbf{R}^+,$$

$$G_E = \frac{1}{-\Delta - (E + i0)}, \quad E \in \mathbf{R} \setminus \{0\},$$

and the integral kernel which goes with it:

$$G_z(x, x') = \frac{i}{4} H_0^{(1)}(\sqrt{z}|x - x'|) = \frac{i}{4} J_0(\sqrt{z}|x - x'|) - \frac{1}{4} Y_0(\sqrt{z}|x - x'|), \quad z \in \mathbf{C} \setminus \mathbf{R}^+. \tag{2.1}$$

These are, respectively, the Hankel and Bessel functions (in the notations of [AS]). Note that $G_z$ is the *free* Green's function, and the interaction will be described purely in terms of the boundary layer operators. Furthermore, the precise form of these functions is not relevant for our purpose, it suffices to know their asymptotic behavior for large and small arguments.

Since we assume that the domain $\Omega$ is connected, we can parameterize the boundary by arclength, by a map $s \mapsto x(s)$ mapping $[0, 2\pi)$ into $\mathbf{R}^2$. Here, we assume without loss of generality that the length of $\Gamma$ is $2\pi$.

We start by defining the "restrictions to the boundary." Let $f$ be a function on $\mathbf{R}^2$. Then

$$(\gamma_\pm f)(s) = \lim_{\varepsilon \downarrow 0} f(x(s) \pm \varepsilon N(s)),$$

$$(\gamma_{N_\pm} f)(s) = \lim_{\varepsilon \downarrow 0} N(s) \cdot (\nabla f)(x(s) \pm \varepsilon N(s)).$$

Here, $N(s)$ denotes the outward normal to $\Gamma$ at $x(s)$ and $\pm$ indicates whether the limit is to be taken (along the normal) from the outside of $\Omega$ (+) or the inside (−). Whenever the direction of the limit is irrelevant, we omit the index $\pm$. In the corners, this definition is problematic, but since we only look at integral kernels, this does not matter. The following identities show where the various restrictions are defined: It is well known [Ne] that for all $\beta > \frac{1}{2}$,

$$\begin{aligned}
\gamma &: H_{\text{loc}}^\beta(\mathbf{R}^2) \to L^2(\Gamma), \\
\gamma^* &: L^2(\Gamma) \to H_{\text{comp}}^{-\beta}(\mathbf{R}^2), \\
\gamma_N &: H_{\text{loc}}^{\beta+1}(\mathbf{R}^2) \to L^2(\Gamma), \\
\gamma_N^* &: L^2(\Gamma) \to H_{\text{comp}}^{-(1+\beta)}(\mathbf{R}^2),
\end{aligned}$$

from which the appropriate domains of $\gamma$ and $\gamma_N$ can be read off. The notation is as follows: Let $\Lambda = (1 - \partial_s^2)^{1/2}$, where $\partial_s$ is the derivative with respect to arclength on $\Gamma$. Then,

$$H^\beta = \{u \in L^2(\Gamma) : \Lambda^\beta u \in L^2(\Gamma)\},$$

for $\beta \geq 0$, $H^\beta_{\text{comp}}$ is the subspace of functions with compact support, and $H^\beta_{\text{loc}}$ denotes functions which are locally in $H^\beta$, see [H]. We also define $H^\beta(\mathbf{R}^2) = \{f \in L^2(\mathbf{R}^2) : (1 - \Delta)^{\beta/2} f \in L^2(\mathbf{R}^2)\}$.

We next define the Single Layer Potential $\Phi_1$ and the Double Layer Potential $\Phi_2$. For $x \in \mathbf{R}^2 \setminus \Gamma$, we have

$$(\Phi_1 u)(x) = \int ds' \, G_z(x, x(s')) u(s'),$$
$$(\Phi_2 u)(x) = \int ds' \, N(s') \cdot (\nabla_2 G_z)(x, x(s')) u(s').$$

The notation $\nabla_2$ means the gradient with respect to the second variable of $G_z$. (The $z$-dependence of $\Phi_j$ is implicit.) A more suggestive notation is

$$\Phi_1 = G_z \gamma^*, \qquad \Phi_2 = G_z \gamma_N^*.$$

Since $G_z$ maps $H^{-\beta}_{\text{comp}}(\mathbf{R}^2)$ to $H^{2-\beta}_{\text{loc}}(\mathbf{R}^2)$ for all $\beta$, we see that

$$\Phi_1 : L^2(\Gamma) \to H^\delta_{\text{loc}}(\mathbf{R}^2), \qquad \Phi_2 : L^2(\Gamma) \to H^{\delta-1}_{\text{loc}}(\mathbf{R}^2),$$

for all $\delta < \frac{3}{2}$.

We finally define

$$\begin{aligned} A_z &= \gamma(G_z \gamma^*), \\ B_z &= \tfrac{1}{2}(\gamma_+ + \gamma_-)(G_z \gamma_N^*), \\ C_z &= \gamma_N(G_z \gamma_N^*). \end{aligned} \qquad (2.2)$$

Henceforth, we will omit the parentheses around $G_z$. It is a well-known fact that the jump discontinuity of the single and double layer potential is 1: More precisely, we have the relation

$$\begin{aligned} 1 &= (\gamma_+ - \gamma_-)(G_z \gamma_N^*), \\ -1 &= (\gamma_{N_+} - \gamma_{N_-})(G_z \gamma^*). \end{aligned} \qquad (2.3)$$

The operators $A_z$, $B_z$, and $C_z$ are defined as maps between the following spaces:

$$\begin{aligned} A_z &: L^2(\Gamma) \to H^1(\Gamma), \\ B_z &: L^2(\Gamma) \to L^2(\Gamma), \\ C_z &: H^1(\Gamma) \to L^2(\Gamma). \end{aligned}$$

For $A_z$, this was shown in [EP1], for $B_z$ we will show it below, and for $C_z$ it will be a consequence of the bounds on $A_z$ and $B_z$.

The following expressions for the integral kernels may make explicit calculations more readable [CH]:

$$A_z(s, s') = (\gamma G_z \gamma^*)(s, s') = G_z(x(s), x(s')), \tag{2.4}$$
$$B_z(s, s') = \sqrt{z} N(s') \cdot \nabla_2 G_z(x(s), x(s')), \tag{2.5}$$
$$B_z^{\mathrm{T}}(s, s') = -\sqrt{z} N(s) \cdot \nabla_1 G_z(x(s), x(s')), \tag{2.6}$$

where $\nabla_j$ is the gradient with respect to the $j^{\mathrm{th}}$ variable. The operator $B^{\mathrm{T}}$ is the transpose of $B$. The case of $C$ is more complicated and will be handled in the Appendix.

Finally, we define the main objects of this paper, namely $\zeta$-functions, one for each of the boundary operators.

**Definition.** We define 4 $\zeta$-functions. We choose some negative number $E_0$ and define, for $z \in \mathbf{C} \setminus \mathbf{R}^+$:

$$\zeta_{\mathbf{DD}}(z) = \det\left(A_{E_0}^{-1} A_z\right),$$
$$\zeta_{\mathbf{DN}}(z) = \det\left((\tfrac{1}{2} - B_{E_0})^{-1}(\tfrac{1}{2} - B_z)\right),$$
$$\zeta_{\mathbf{ND}}(z) = \det\left((\tfrac{1}{2} + B_{E_0})^{-1}(\tfrac{1}{2} + B_z)\right),$$
$$\zeta_{\mathbf{NN}}(z) = \det\left(C_{E_0}^{-1} C_z\right).$$

The first index will refer to the interior boundary condition and the second to the exterior boundary condition.

**Remark.** There is a close relationship among the 4 $\zeta$-functions, which is a consequence of the identity:

$$C_z = (\tfrac{1}{2} + B_z^{\mathrm{T}}) A_z^{-1} (\tfrac{1}{2} - B_z) = (\tfrac{1}{2} - B_z^{\mathrm{T}}) A_z^{-1} (\tfrac{1}{2} + B_z). \tag{2.7}$$

The identity Eq.(2.7) follows from the following considerations: Fix $u$ and define the single layer potential $\Phi = \Phi_1 u$. Then we have $\gamma_- \Phi = Au$ and

$$\gamma_{N_-} \Phi = (\tfrac{1}{2} + B^{\mathrm{T}}) u.$$

We can write the same function $\Phi$ as a double layer potential: if $\Phi = \Phi_2 v$ inside $\Omega$, then $\gamma_- \Phi = (\tfrac{1}{2} - B)v$ and

$$\gamma_{N_-} \Phi = Cv.$$

Hence we find

$$Au = (\tfrac{1}{2} - B)v, \quad (\tfrac{1}{2} + B^{\mathrm{T}})u = Cv,$$

from which the first identity in Eq.(2.7) follows. The second identity can be obtained by repeating the above arguments for $\Omega^c$ in place of $\Omega$, i.e., the limits are taken from the outside.

Using the identity Eq.(2.7), it is almost obvious that it suffices to study the $\zeta$-functions $\zeta_{\mathbf{DD}}$, $\zeta_{\mathbf{DN}}$, and $\zeta_{\mathbf{ND}}$. Then $\zeta_{\mathbf{NN}}$ can be expressed as

$$\begin{aligned}
\zeta_{\mathbf{NN}}(z) &= \det\left(C_{E_0}^{-1} C_z\right) \\
&= \det\left((\tfrac{1}{2} - B_{E_0})^{-1} A_{E_0} (\tfrac{1}{2} + B_{E_0}^{\mathrm{T}})^{-1} (\tfrac{1}{2} + B_z^{\mathrm{T}}) A_z^{-1} (\tfrac{1}{2} - B_z)\right) \\
&= \det\left((\tfrac{1}{2} + B_{E_0}^{\mathrm{T}})^{-1} (\tfrac{1}{2} + B_z^{\mathrm{T}}) A_z^{-1} (\tfrac{1}{2} - B_z)(\tfrac{1}{2} - B_{E_0})^{-1} A_{E_0}\right) \\
&= \zeta_{\mathbf{ND}}(z) \cdot \det\left(A_z^{-1} (\tfrac{1}{2} - B_z)(\tfrac{1}{2} - B_{E_0})^{-1} A_{E_0}\right) \\
&= \zeta_{\mathbf{ND}}(z) \cdot \det\left(A_{E_0} A_z^{-1} (\tfrac{1}{2} - B_z)(\tfrac{1}{2} - B_{E_0})^{-1}\right) \\
&= \zeta_{\mathbf{ND}}(z) \cdot \det\left(A_{E_0} A_z^{-1}\right) \cdot \det\left((\tfrac{1}{2} - B_z)(\tfrac{1}{2} - B_{E_0})^{-1}\right) \\
&= \zeta_{\mathbf{ND}}(z) \cdot \zeta_{\mathbf{DD}}^{-1}(z) \cdot \zeta_{\mathbf{DN}}(z)\,.
\end{aligned}$$

This identity allows to avoid the use of $C$ which is more complicated to compute than $A$ or $B$.

## 3. The Relation Between the Scattering Phase and the Density of States

Our main result is the following set of identities:

**Theorem 3.1.** *For any choice of $\mathbf{B}, \mathbf{B}' \in \{\mathbf{D}, \mathbf{N}\}$, the total scattering phases, the integrated density of states and the $\zeta$ functions are related (for $E > 0$), by*

$$\pi N_{\mathbf{B}}(E) = \Theta_{\mathbf{B}'}(E) - \operatorname{Im} \log \zeta_{\mathbf{B},\mathbf{B}'}(E + i0)\,. \tag{3.1}$$

**Remark.** The relation Eq.(2.7) is reflected through the Eqs.(3.1), since the 4 possible left hand sides are linearly dependent. We furthermore have the bounds:

**Theorem 3.2.** *One has the following bounds, valid for $E > 2$:*

$$\begin{aligned}
0 &\leq \Theta_{\mathbf{D}}(E) - \pi N_{\mathbf{D}}(E) \leq \mathrm{const.}\, \mathrm{E}^{1/2} \log \mathrm{E}\,, &\text{(3.2)} \\
&|\Theta_{\mathbf{N}}(E) - \pi N_{\mathbf{D}}(E)| \leq \mathrm{const.}\, \mathrm{E}^{1/2} \log \mathrm{E}\,, &\text{(3.3)} \\
&|\Theta_{\mathbf{D}}(E) - \pi N_{\mathbf{N}}(E)| \leq \mathrm{const.}\, \mathrm{E}^{1/2} \log \mathrm{E}\,, &\text{(3.4)} \\
0 &\leq \pi N_{\mathbf{N}}(E) - \Theta_{\mathbf{N}}(E) \leq \mathrm{const.}\, \mathrm{E}^{1/2} \log \mathrm{E}\,. &\text{(3.5)}
\end{aligned}$$

**Remark.** With slightly more complicated expressions due to threshold effects the formulas above extend to $E = 0$.

**Discussion.** The above results describe a close relation between the *integrated* density of states and *total* scattering phase. Thus, they are much weaker than the spectral duality result ("inside-outside duality"), conjectured in [DS] and proved in [EP1], but they generalize and extend the pioneering result of [JK]. To complete the picture, we state here the result which relates *individual* eigenvalues and eigenphases:

**Theorem 3.3.** *If $\Omega$ is a standard domain, then $E^*$ is an eigenvalue of $-\Delta_{\Omega,\mathbf{D}}$ of multiplicity $m$ if and only $m$ eigenphases of the S-matrix in $\Omega^c$ with Dirichlet boundary conditions approach $\pi$ from below as $E \uparrow E^*$.*
*If $\Omega$ is a standard domain, then $E^*$ is an eigenvalue of $-\Delta_{\Omega,\mathbf{N}}$ of multiplicity $m$ if and only $m$ eigenphases of the S-matrix in $\Omega^c$ with Neumann boundary conditions approach $0$ from above as $E \downarrow E^*$.*

**Remark.** The first part was shown in [EP1], the second part is new. We do not expect any similar result for the case when the boundary condition for the inside and the outside problem are not the same. Our convention of scattering phase is that the eigenphases of the (unitary) S-matrix are $\exp(-2i\theta_\ell(E))$, $\ell = 1, 2, \ldots$.

**Remark.** We next wish to comment on the bounds in Theorem 3.2 and their possible optimality. The growth of $\mathrm{Im}\log\zeta$, has, in our view, two different origins in the case of $\zeta_{\mathbf{DD}}$ and $\zeta_{\mathbf{NN}}$ when compared to the "mixed" cases $\zeta_{\mathbf{DN}}$ and $\zeta_{\mathbf{ND}}$. In the case of $\zeta_{\mathbf{DD}}$ the growth can be traced back to the different Weyl asymptotics of $\Theta_{\mathbf{D}}$ and $N_{\mathbf{D}}$ [SU], namely

$$\pi N_{\mathbf{D}}^{\mathbf{W}}(E) = \frac{|\Omega|}{4}E - \frac{|\partial\Omega|}{4}E^{1/2} + \mathcal{O}(1),$$

$$\Theta_{\mathbf{D}}(E) = \frac{|\Omega|}{4}E + \frac{|\partial\Omega|}{4}E^{1/2} + \mathcal{O}(1),$$

where the superscript $\mathbf{W}$ indicates the Weyl approximation, so that already at the "average" level the two quantities differ by $|\partial\Omega|E^{1/2}$. A similar formula holds for the pure Neumann case. On the other hand, in the case of $\zeta_{\mathbf{ND}}$, there is a different asymptotics since

$$\pi N_{\mathbf{N}}^{\mathbf{W}}(E) = \frac{|\Omega|}{4}E + \frac{|\partial\Omega|}{4}E^{1/2} + \mathcal{O}(1),$$

and therefore there is a *cancellation* of the terms of order $E^{1/2}$ at the Weyl level.

We next discuss in detail the question whether this cancellation implies better bounds in Eqs.(3.3) and (3.4). (We neglect here the issue of eliminating the factor $\log E$.)

A *first* possibility might seem a proof using the properties of $B$. Indeed, in the integral kernel of $B$ there appears the product $N(s') \cdot (x(s) - x(s'))/|x(s) - x(s')|$ which goes to 0 as $s' \to s$, so that in the detailed bounds one additional order cancels when compared to the bound on $\partial_s A(s, s')$, which occurs in the estimate for $\mathrm{Im}\log\zeta_{\mathbf{DD}}$ (see Eq.(5.10)). However, we still cannot exclude that the bounds in Eqs.(3.3) and (3.4) are optimal, since actually the cancellation does *not* take place at intermediate distances, i.e., $|s - s'| = \mathcal{O}(1)$.

A *second* possibility is provided by the very detailed results from the methods of pseudo-differential operators. The following discussion is a summary of the papers by Seeley, Melrose, Ivrii, Buslaev, Robert, and Vasil'ev, Safarov[VS]. The major new ingredient here is the notion of the set of periodic orbits of same length. We say that a billiard has property S (for synchronous) if there are "many" periodic orbits in the following sense. Consider a periodic orbit, of period $T$. Let $\varphi^T(z)$ denote the phase space point reached from $z \in \Omega \times S^1$ (initial position and initial direction of the orbit) after time $T$, i.e., the end point of the billiard trajectory (in phase space) with initial time $T$. Let $z^*$ be the periodic point: $\varphi^T(z^*) = z^*$. We say this orbit is "absolutely periodic" [VS] if

$$f(z) \equiv |\varphi^T(z) - z|$$

has a zero of infinite order at $z = z^*$. In other words, the returning rays have infinite focusing in a neighborhood of $z^*$. A billiard has property S if the set of absolutely periodic points has positive measure in phase space. (For example, if $f(z)$ is a "devil's staircase," then it has a set of full measure of points where $f(z) - f(z^*)$ vanishes of infinite order when $z \to z^*$.) Examples of billiards with property S are given in [VS]. If a billiard does not have property S, we say that it has property A (for asynchronous). When talking about scattering, this condition is to be applied to *exterior* orbits, which might for example be trapped in the "outside" of an obstacle. If a billiard has property S, the term of order $E^{1/2}$ in the Weyl series is modified by an oscillating amplitude, $W(E)$, which can in principle be computed from the knowledge of the synchronous set. If $\Omega$ is convex and the boundary is an analytic curve, then one has property A, but in most other cases, it is difficult to decide whether a domain has property A or S. The following table summarizes the known results for $\mathcal{C}^\infty$ boundary (in dimension 2, in odd dimensions, slightly more is known). The upper sign is for Dirichlet, the lower for Neumann boundary conditions.

| Property | $\pi N(E)$ | $\Theta(E)$ |
|---|---|---|
| A | $\frac{|\Omega|}{4}E \mp \frac{|\partial\Omega|}{4}E^{1/2} + o(E^{1/2})$ | $\frac{|\Omega|}{4}E \pm \frac{|\partial\Omega|}{4}E^{1/2} + o(E^{1/2})$ |
| S | $\frac{|\Omega|}{4}E \mp \frac{|\partial\Omega|}{4}W(E)E^{1/2} + o(E^{1/2})$ | $\frac{|\Omega|}{4}E + \mathcal{O}(E^{1/2})$ |

If the boundary is Lipshitz then it is only known that $\Theta(E) = |\Omega|E/4 + o(E)$, in all cases [R1, R2]. Applying the results of this table to our questions, we see that the only known improvement over our bound seems to be a bound of $o(E^{1/2})$ in Eqs.(3.3) and (3.4) when property A holds. The disc is an example of this case [SU].

A *third* way to view these problems is in the context of scattering resonances [He], although we have no rigorous results to offer. Consider domains with trapped orbits in $\Omega^c$. Then we expect resonances and these may contribute to the growth of the $\zeta$-function. To decide how much they grow, one would have to know if these resonances stay near the real axis. If they do, they will contribute a term $\mathcal{O}(E^{1/2})$ to $\operatorname{Im} \log \zeta_{\mathbf{DN}}$. But if they move away fast enough from the real axis, as the energy increases, they might as well only contribute $\mathcal{O}(1)$. In that case, the bound Eq.(3.3) would not be optimal.

## 4. The Krein Formula

Using the methods of [EP2], one can derive from the identities of Theorem 3.1 a corresponding set of Krein trace formulas and, in the case of identical boundary conditions only, a Green's formula. We shall state them here without proof.

**Definition.** We let $H_0 = -\Delta$, and we define the "interacting" Hamiltonians

$$H_{\mathbf{B},\mathbf{B}'} = -\Delta_{\Omega,\mathbf{B}} \oplus -\Delta_{\Omega^c,\mathbf{B}'} .$$

Furthermore, we denote by $\lambda_{j,\mathbf{B}}$ the eigenvalues (counted with multiplicity) of $-\Delta_{\Omega,\mathbf{B}}$, and by $S_{E,\mathbf{B}'}$ the on-shell S-matrix for scattering on $\Omega^c$, with boundary conditions $\mathbf{B}'$. With these notations, one has the

**Theorem 4.1.** *For every choice of* $\mathbf{B}, \mathbf{B}' \in \{\mathbf{D}, \mathbf{N}\}$ *and for all* $F \in \mathcal{S}(\mathbf{R}^+)$ *with support in* $\{E : E > 0\}$, *one has the identity*

$$\begin{aligned}\operatorname{Tr}\bigl(F(H_{\mathbf{B},\mathbf{B}'}) - F(H_0)\bigr) &= \sum_n F(\lambda_{j,\mathbf{B}}) \\ &\quad + \frac{1}{\pi i} \int dE\, F(E) \operatorname{Tr}\bigl(S^*_{E,\mathbf{B}'} \partial_E S_{E,\mathbf{B}'}\bigr) .\end{aligned} \quad (4.1)$$

The proof follows very closely the one in [EP2, Section 4], and will be omitted.

We next state some identities for the S-matrix:

**Definition.** We start by defining the operator $\Sigma_E$, which is "restriction to the energy surface $E$." Let $p \in \mathbf{R}^2$, $p \cdot p = E$. Then

$$\begin{aligned}(\Sigma_E \psi)(p) &= \int_{\mathbf{R}^2} d^2y\, e^{-ip\cdot y} \psi(y) , \\ (\Sigma_E^* \chi)(x) &= \int_0^{2\pi} d\varphi\, e^{ip(\varphi)x} \chi(p(\varphi)) ,\end{aligned}$$

where $p(\varphi) = \sqrt{E}(\cos\varphi, \sin\varphi)$. If we denote by $F_E$ the energy surface, $F_E = \{p \in \mathbf{R}^2 : p \cdot p = E\}$, then we see that for all $\beta \geq 0$,

$$\begin{aligned}\Sigma_E^* &: L^2(F_E) \to H^\beta_{\mathrm{loc}}(\mathbf{R}^2) , \\ \Sigma_E &: H^{-\beta}_{\mathrm{comp}}(\mathbf{R}^2) \to L^2(F_E) .\end{aligned}$$

**Theorem 4.2.** *The S-matrix at energy $E$ can be expressed as follows:*

$$\begin{aligned}S_{\mathbf{D}}(E) &= 1 - 2\pi i \Sigma_E \gamma^* A^{-1}_{E+i0} \gamma \Sigma_E^* \\ &= 1 - 2\pi i \Sigma_E \gamma^*_{N_+} \bigl(\tfrac{1}{2} + B^{\mathrm{T}}_{E+i0}\bigr)^{-1} \gamma \Sigma_E^* . \\ S_{\mathbf{N}}(E) &= 1 - 2\pi i \Sigma_E \gamma^*_{N_+} C^{-1}_{E+i0} \gamma_{N_+} \Sigma_E^* \\ &= 1 - 2\pi i \Sigma_E \gamma^* \bigl(\tfrac{1}{2} - B_{E+i0}\bigr)^{-1} \gamma_{N_+} \Sigma_E^* .\end{aligned}$$

Note that $\gamma \Sigma_E^*$ and all the other combinations of $\gamma$ and $\Sigma$ used above are well-defined.

We finally state two identities between Green's functions. Note that no such identity is available in the case when the inside boundary condition is not the same as the outside boundary condition.

**Theorem 4.3.** *The Green's functions satisfy the following identities:*

$$G_{\mathbf{DD}} = G - G\gamma^* A^{-1} \gamma G ,$$
$$G_{\mathbf{NN}} = G - G\gamma_N^* C^{-1} \gamma_N G .$$

The boundary layer integral lends itself in a very natural and systematic way for the computation of $\zeta_{\mathbf{DN}}$, as well as for a determination of the eigenvalues and, to some extent of the scattering phases for the Dirichlet, resp. the Neumann problem. These algorithms work, at present, only for the case of domains where $\Gamma$ is $\mathcal{C}^1$, i.e., corners are excluded, but jumps in the second derivative are allowed. Consider the integral kernel $B(s, s')$. To discretize it, we choose an ordered sequence of points $s_j$ on the boundary (this method is also used in [HS]). If the boundary is smooth, it is advisable to choose the points equidistant in arclength, since then the method is of infinite order in the step size [Ha]. In the other cases, we have chosen unequal steps, and in particular 2 points at distance 0 at every discontinuity of the second derivative of the boundary, one point as the limit on either side.

## 5. Bounds on the Double Layer Potential

Our main ingredient for the proof of all the results stated so far are *Structure Theorems*, which describe the detailed regularity properties of the operators $A_z$ and $B_z$, and hence, by Eq.(2.7), also those of $C_z$. We first state this result:

**Definitions.** If $C$ is a compact operator, we let $s_n(C)$, $n = 1, 2, \ldots$ denote the eigenvalues of $(C^*C)^{1/2}$ in decreasing order. One defines the weak Schatten classes (for $1 \leq p < \infty$), as the set of those $C$ for which

$$\langle C \rangle_p = \sup_n n^{1/p} s_n(C) ,$$

is finite. We also define the associated norms

$$\|C\|_p = \left( \sum_{n=1}^{\infty} s_n(C)^p \right)^{1/p} .$$

We let $P_0$ denote the orthogonal projection onto the constant functions in $L^2(\Gamma)$. Recall also that $\Lambda = (1 - \partial_s^2)^{1/2}$.

**Definitions.** We need some precisions concerning the branch cuts in the definition of $G_z$, cf. Eq.(2.1). Let $\mathcal{E}$ denote $\{z : z \in \mathbf{C}, z \notin \mathbf{R}^+\}$, and let $\mathcal{R}$ denote the Riemann surface associated with the logarithm. The function $H_0^{(1)}$ is defined on $\mathcal{R}$, and the integral kernel $G_z$ is defined for $z \in \mathcal{E}$, with the convention that $\text{Im } z^{1/2} > 0$ for $z \in \mathcal{E}$. Then we have

**Structure Theorem 5.1.**
– For $z \in \mathcal{R}$ the operator $\Lambda A_z$ has the following representation:

$$\Lambda A_z = \tfrac{1}{2} + Q_A + R_A + T_z^{(A)}, \tag{5.1}$$

where $Q_A$ is bounded and of norm $\|Q_A\| < \tfrac{1}{2}$, where $R_A$ is Hilbert-Schmidt, and where $T_z^{(A)}$ is trace class. Moreover, there is a constant $K$ so that for $z \in \mathcal{E}$, one has the bounds

$$\begin{aligned} \langle T_z^{(A)} + \tfrac{1}{2} P_0 \log z \rangle_{2/3} &\leq K|z|^{3/4} |\log z|, \\ \|T_z^{(A)} \Lambda + \tfrac{1}{2} P_0 \log z\|_2 &\leq K|z|^{3/4} |\log z|. \end{aligned} \tag{5.2}$$

The operators $Q_A$ and $R_A$ do not depend on $z$.

– For $z \in \mathcal{R}$ the operator $B_z$ has the following representation:

$$B_z = Q_B + R_B + T_z^{(B)}, \tag{5.3}$$

where $Q_B$ is bounded and of norm $\|Q_B\| < \tfrac{1}{2}$, where $R_B$ is Hilbert-Schmidt, and where $T_z^{(B)}$ is trace class. Moreover, there is a constant $K$ so that for $z \in \mathcal{E}$, one has the bounds

$$\begin{aligned} \langle T_z^{(B)} \rangle_{2/3} &\leq K|z|^{3/4} |\log z|, \\ \|T_z^{(B)} \Lambda\|_2 &\leq K|z|^{3/4} |\log z|. \end{aligned} \tag{5.4}$$

The operators $Q_B$ and $R_B$ do not depend on $z$.

**Remarks.** Note that $B_z$ already "contains" a derivative (the normal derivative), whereas in $A_z$ the derivative is provided by $\Lambda$. Note also that while the operator $\tfrac{1}{2}$ seems to be absent from $B_z$, it reappears naturally through the very definition of $B_z$, Eqs.(2.2) and (2.3). So the results for $A_z$ and $B_z$ are in fact quite similar. In particular, as discussed before, we do not get better results for $T_z^{(B)}$ than for $T_z^{(A)}$, although such a result might have been expected from a local analysis.

**Proof.** The proofs for the operators $A_z$ have been given in [EP2], [EP1], so we deal here only with $B_z$.

We note first that the Green's function equals $G_z(x, x') = (i/4) H_0^{(1)}(z^{1/2} |x - x'|)$, where $H_0^{(1)}$ is the Hankel function, cf. [AS, 9.1],

$$H_0^{(1)}(w) = J_0(w) + i Y_0(w).$$

For $z \in \mathcal{E}$ (which is really a complex energy) we let $k = z^{1/2}$. Starting from Eq.(2.5), we see that the integral kernel for $B_z$ is

$$B_z(s, s') = -k \frac{N(s') \cdot (x(s) - x(s'))}{|x(s) - x(s')|} G'(k|x(s) - x(s')|), \tag{5.5}$$

where $G(x) = \frac{i}{4}H_0^{(1)}(x)$. To study $B_z$, we start with Eq.(5.5). First observe that

$$\partial_z H_0^{(1)}(z) = -(J_1(z) + iY_1(z)) = -H_1^{(1)}(z).$$

Substituting the definition of $G$, we get

$$B_z(s,s') = \frac{ik}{4}H_1^{(1)}(k|x(s) - x(s')|)\frac{N(s') \cdot (x(s) - x(s'))}{|x(s) - x(s')|}.$$

We introduce the notations

$$B_z(s,s') = \frac{ik}{4}N(s') \cdot D(s,s')H_1^{(1)}(kr),$$

with

$$r(s,s') = |x(s) - x(s')|,$$
$$D(s,s') = (x(s) - x(s'))/r(s,s').$$

The local behavior of $H_1^{(1)}$ is given by

$$H_1^{(1)}(r) = -\frac{2i}{\pi r} + \mathcal{O}(r(1 + \log r)),$$

and we introduce the regular part $g$ of $H_1^{(1)}$:

$$\begin{aligned}
g(r) &\equiv H_1^{(1)}(r) + \frac{2i}{\pi r} \\
&= \frac{2i}{\pi}J_1(r)\log r + \mathcal{O}(r) \\
&= \frac{i}{\pi}r\log r + \mathcal{O}(r),
\end{aligned} \qquad (5.6)$$

cf. [AS, 9.1]. We will also need in the sequel the representation

$$g'(r) = \frac{i}{\pi}\log r + \mathcal{O}(1) = \frac{g(r)}{r} + \mathcal{O}(1).$$

With these notations, we define the regular and singular parts of $B_z$:

$$\begin{aligned}
B^{\text{sing}}(s,s') &= \frac{1}{2\pi r(s,s')}N(s') \cdot D(s,s'), \\
B_z^{\text{reg}}(s,s') &= \frac{ik}{4}N(s') \cdot D(s,s')g(kr(s,s')),
\end{aligned}$$

so that $B_z = B^{\text{sing}} + B_z^{\text{reg}}$. Note that $B^{\text{sing}}$ does *not* depend on $z$.

**Proposition 5.2.** *The operator $B_z^{\text{reg}}$ is in the Schatten class 2/3 and for all $z \in \mathcal{E}$ one has the bound*

$$\langle B_z^{\text{reg}} \rangle_{2/3} \leq \text{const.} |z|^{3/4}(1 + |\log z|) . \tag{5.7}$$

**Proof.** In order to bound $B_z^{\text{reg}}$, we consider the quantity $\partial_s B_z^{\text{reg}}(s, s')$. From Eq.(5.6), we get the representation

$$g'(r) = \frac{i}{\pi} \log r + \mathcal{O}(1) = \frac{g(r)}{r} + \mathcal{O}(1) .$$

Furthermore, we have, with $T(s) = \partial_s x(s)$,

$$\partial_s r(s, s') = T(s) \cdot D(s, s') ,$$
$$\partial_s D(s, s') = \frac{1}{r(s, s')} \left( T(s) - D(s, s')(T(s) \cdot D(s, s')) \right) ,$$

and therefore

$$\begin{aligned}
\partial_s B_z^{\text{reg}}(s, s') &= \frac{ik}{4} N(s') \frac{1}{r} \left( (T - D(T \cdot D))g(kr) + Dg'(kr)krT \cdot D \right) \\
&= \frac{ik}{4} N(s') \frac{1}{r} \left( (T - D(T \cdot D))g(kr) + D\left(\frac{g(kr)}{kr} + \mathcal{O}(1)\right)krT \cdot D \right) \\
&= \frac{ik}{4} N(s') \frac{1}{r} \left( T(s)g(kr(s, s')) + kr\mathcal{O}(1) \right) \\
&= k^2 \left( \frac{i}{4} N(s') \cdot T(s) \left( \frac{i}{\pi} \log(kr(s, s')) + \mathcal{O}(1) \right) + \mathcal{O}(1) \right) \\
&= -\frac{k^2}{4\pi} \left( T(s) \cdot N(s') \log(kr(s, s')) + \mathcal{O}((kr)^0) \right) .
\end{aligned}$$

We next bound the Hilbert-Schmidt norm of $\partial_s B_z^{\text{reg}}$ and of $B_z^{\text{reg}}$.

**Lemma 5.3.** *There is a $K < \infty$ such that for all $|z| > 2$, one has the bounds*

$$\|\partial_s B_z^{\text{reg}}\|_2 \leq K|z|^{3/4}|\log z| , \tag{5.8}$$
$$\|B_z^{\text{reg}}\|_2 \leq K|z|^{3/4}|\log z| . \tag{5.9}$$

**Proof.** We only show Eq.(5.8) and leave the proof of Eq.(5.9) to the reader. We split the integral

$$I \equiv \int ds\, ds' \left|\partial_s B_z^{\text{reg}}(s, s')\right|^2$$

into two parts, $I = I_< + I_>$ corresponding to the integration region $kr(s, s') < 1$ and its complement. We then get the bounds, observing that $s$ and $s'$ are bounded:

$$\begin{aligned}
I_< &= \mathcal{O}(k^4) \int_{kr<1} \left|1 + |\log(kr)|\right|^2 \\
&\leq \mathcal{O}(k^4) k^{-1} \int_0^1 dx\, (1 + |\log x|)^2 \leq \mathcal{O}(k^3) .
\end{aligned} \tag{5.10}$$

We have used here that $kr(s,s') < 1$ implies $|s-s'| < \mathcal{O}(k^{-1})$. In the complement, we use that the integrand is bounded by $k^4(\log kr)^2$, and a scaling argument shows that $I_> \leq \mathcal{O}(k^3(\log k)^2)$, as $k \to \infty$. The proof of Eq.(5.8) is complete.

**Proof of Proposition 5.2.** Note first that

$$\Lambda \;=\; (1 + (i\partial_s)^2)^{1/2} \;=\; |1 + \partial_s| \;=\; U(1 + \partial_s)\,,$$

where $U$ is unitary. Therefore, the Schwarz inequality and Eqs.(5.8), (5.9), imply

$$\|\Lambda B_z^{\mathrm{reg}}\|_2 \;\leq\; K|z|^{3/4}|\log z|\,.$$

Since the spectrum of $\Lambda$ is $\{\sqrt{1+n^2}\}_{n\in\mathbf{Z}}$, we get $\langle \Lambda^{-1}\rangle_1 < \infty$. We next use the inequality

$$\langle C_1 C_2\rangle_p \;\leq\; 2^{1/p}\langle C_1\rangle_q \langle C_2\rangle_r\,,$$

valid for all $q > 0$, $r > 0$, $p^{-1} = q^{-1} + r^{-1}$, see [BS]. This implies

$$\langle B_z^{\mathrm{reg}}\rangle_{3/2} \;=\; \langle \Lambda^{-1}\Lambda B_z^{\mathrm{reg}}\rangle_{3/2} \;\leq\; 2^{3/2}\langle \Lambda B_z^{\mathrm{reg}}\rangle_2 \langle \Lambda^{-1}\rangle_1 \;\leq\; \mathrm{const.}\,\|\Lambda B_z^{\mathrm{reg}}\|_2\,. \qquad (5.11)$$

The proof of Proposition 5.2 is complete.

We consider next the operator $B^{\mathrm{sing}}$. Recall again that it does *not* depend on $z$. Our result here is

**Proposition 5.4.** *The operator $B^{\mathrm{sing}}$ has the following representation:*

$$B^{\mathrm{sing}} \;=\; Q_B + R_B\,,$$

*where $Q_B$ is bounded and of norm $\|Q_B\| < \tfrac{1}{2}$, and $R_B$ is Hilbert-Schmidt.*

**Proof of the Structure Theorem 5.1.** This is an immediate consequence of Proposition 5.4 and Proposition 5.2 and an inequality of the type of Eq.(5.11).

**Remark.** We shall use the notation $\approx$ to denote equality modulo Hilbert-Schmidt operators, i.e., $B_1(s,s') \approx B_2(s,s')$ means that $B_1 - B_2$ is Hilbert-Schmidt.

**Proof of Proposition 5.4.** We shall localize the integral kernel of $B^{\mathrm{sing}}$ near the corners and in the complement of this region, and give a different treatment for the two pieces. To localize, assume the boundary has exactly $n$ corners, and define $n$ smooth functions $\chi_j^{(0)} : \Gamma \times \Gamma \to \mathbf{R}$ which are equal to 1 when $s, s'$ are both close to the corner $j$, and such that they have disjoint supports. We then further decompose into the $2n$ functions

$$\chi_j^{\mathrm{opposite}}(s,s') \;=\; \chi_j^{(0)}(s,s') \cdot \chi_{j,0}^{\mathrm{opposite}}(s,s')\,,$$
$$\chi_j^{\mathrm{same}}(s,s') \;=\; \chi_j^{(0)}(s,s') \cdot \chi_{j,0}^{\mathrm{same}}(s,s')\,,$$

where $\chi_{j,0}^{\text{opposite}}$ is 1 if $s$, $s'$ are on opposite sides of (and close to) corner $j$, and 0 otherwise, and $\chi_{j,0}^{\text{same}} = 1 - \chi_{j,0}^{\text{opposite}}$. We next describe the local behavior of $B^{\text{sing}}$ on a smooth part of $\Gamma$ and near a corner of $\Gamma$.

**Lemma 5.5.** *On the smooth pieces of $\Gamma$ one has*

$$B^{\text{sing}}(s, s') \approx -\frac{1}{4\pi}\varkappa(s) \,. \tag{5.12}$$

*Here, $\varkappa(s)$ is the curvature at $s$.*

**Remark.** These approximations hold near the corners as well if $s$, $s'$ are on the same side of the corner. In particular,

$$B^{\text{sing}}(s, s')\left(1 - \sum_{j=1}^{n} \chi_j^{\text{opposite}}\right)(s, s')$$

$$\approx -\frac{1}{4\pi}\varkappa(s)\left(1 - \sum_{j=1}^{n} \chi_j^{\text{opposite}}\right)(s, s') + \mathcal{O}(s - s') \,.$$

From this and the compactness of $\Gamma$ it follows that the corresponding contribution of $B^{\text{sing}}$ is Hilbert-Schmidt.

**Proof of Lemma 5.5.** We recall the definition of $B^{\text{sing}}$:

$$B^{\text{sing}}(s, s') = \frac{1}{2\pi}\frac{N(s') \cdot (x(s) - x(s'))}{|x(s) - x(s')|^2} \,. \tag{5.13}$$

Since we consider the smooth part of $\Gamma$, we get, with $T(s)$ denoting the tangent vector,

$$\begin{aligned} x(s) - x(s') &= (s - s')T(s') - \tfrac{1}{2}(s - s')^2\varkappa(s')N(s') + o\big((s - s')^2\big) \,, \\ |x(s) - x(s')|^2 &= |s - s'|^2\big(1 + o\big((s - s')^2\big)\big) \,, \\ |x(s) - x(s')| &= |s - s'|\big(1 + o\big((s - s')^2\big)\big) \,, \\ N(s') \cdot \big(x(s) - x(s')\big) &= -\tfrac{1}{2}(s - s')^2\varkappa(s') + o\big((s - s')^2\big) \,. \end{aligned}$$

Substituting in Eq.(5.13), the assertion Lemma 5.5 follows.

We next study the behavior "across" one corner, i.e., $B^{\text{sing}}\chi_j^{\text{opposite}}$.

**Lemma 5.6.** *Modulo Hilbert-Schmidt operators, the operators $B^{\text{sing}}\chi_j^{\text{opposite}}$ are bounded in norm by $\frac{1}{2}|\cos(\frac{\alpha_j}{2})|$, where $\alpha_j$ is the interior angle at the corner $j$. Since the $B^{\text{sing}}\chi_j^{\text{opposite}}$ have disjoint supports, it follows that there is a Hilbert-Schmidt operator $K$ (which is independent of the energy $z$), for which one has the inequality*

$$\left\|B^{\text{sing}}\sum_{j=1}^{n}\chi_j^{\text{opposite}} - K\right\| \leq \max_{j=1,\ldots,n}\tfrac{1}{2}|\cos(\tfrac{\alpha_j}{2})| \,. \tag{5.14}$$

**Proof.** We assume that the boundary is locally straight. We leave to the reader the study of the Hilbert-Schmidt correction terms generated by curvature near a corner, see [EP1]. Assume for convenience that the corner is at the origin and use coordinates $s$ and $t = -s'$ near the corner, when $s > 0$ and $s' < 0$.

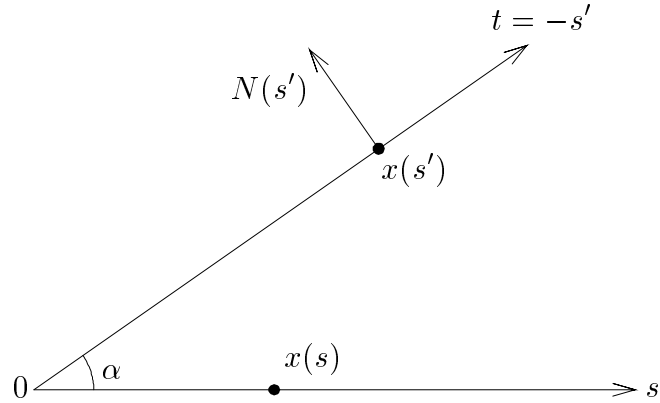

**Fig. 1**: The local coordinate system.

Elementary geometry leads to

$$|x(s) - x(s')|^2 = s^2 + t^2 - 2st\cos\alpha ,$$
$$N(s') \cdot (x(s) - x(s')) = N(s')x(s) = -s\sin\alpha .$$

Redoing this calculation for $s < 0$, $s' > 0$, we find that for all $s$, $s'$ satisfying $ss' < 0$ one has

$$N(s') \cdot (x(s) - x(s')) = -|s|\sin\alpha .$$

We shall consider $Q = B^{\text{sing}}\chi^{\text{opposite}}$ near one corner. We find, that $Q$ equals (locally)

$$Q(s, s') = \chi(ss' < 0)\frac{1}{2\pi}\frac{-|s|\sin\alpha}{s^2 + s'^2 + 2ss'\cos\alpha}$$
$$= \chi(ss' < 0)\frac{-\operatorname{sign} s}{4\pi i}\left(\frac{e^{i\alpha}}{s + e^{i\alpha}s'} - \frac{e^{-i\alpha}}{s + e^{-i\alpha}s'}\right) .$$

We study, as in [EP1], the operator $Q$ acting on $L^2(\mathbf{R})$. This is done by decomposing first $L^2(\mathbf{R}) = L^2(\mathbf{R}^+) \oplus L^2(\mathbf{R}^+)$, using the map $u(s) \mapsto (u_+(s), u_-(s))$ with

$$u(s) = \begin{cases} u_+(s), & \text{when } s > 0, \\ u_-(-s), & \text{when } s < 0. \end{cases}$$

Having gone to unbounded coordinates, we can use them for an explicit calculation.

The operator $Q$ is diagonalized by the Mellin transformation $\mathcal{M}$, defined by

$$(\mathcal{M}f)(\lambda) = \frac{1}{\sqrt{\pi}} \int_0^\infty ds\, s^{i\lambda - 1/2} f(s)\,,$$

as we shall show now. Indeed, this is intuitively clear since $\mathcal{M}$ diagonalizes dilatations. Note that $\mathcal{M} : L^2(\mathbf{R}^+) \to L^2(\mathbf{R})$ is unitary. With the above notation, we see that $\mathcal{M}Qu$ is given by

$$(\mathcal{M}Qu)_\sigma(\lambda)$$
$$= -\frac{1}{\sqrt{\pi}} \int_0^\infty ds\, s^{i\lambda - 1/2} \int_0^\infty \frac{ds'}{2\pi i} \left(\tfrac{1}{2} \frac{1}{s - s'e^{i\alpha}} - \tfrac{1}{2} \frac{1}{s - s'e^{-i\alpha}}\right) u_{-\sigma}(s')\,,$$

where $\sigma \in \{+, -\}$. Replacing the integration variable $s$ by $ss'$ and noting that the integrand is homogeneous of degree $i\lambda + 1/2$ in $s'$, we get

$$(\mathcal{M}Qu)_\sigma(\lambda) = -\int_0^\infty \frac{ds}{2\pi i} s^{i\lambda - 1/2} \left(\tfrac{1}{2} \frac{1}{s - e^{i\alpha}} - \tfrac{1}{2} \frac{1}{s - e^{-i\alpha}}\right) (\mathcal{M}u_{-\sigma})(\lambda)\,,$$
$$\equiv c(\lambda)(\mathcal{M}u_{-\sigma})(\lambda)\,.$$

Thus, $Q$ becomes matrix multiplication under the Mellin transform. We next evaluate the integral $c(\lambda)$. Note that the integrand is $\mathcal{O}(s^{-3/2})$ at infinity and $\mathcal{O}(s^{-1/2})$ near 0. Therefore, for large $R$, we find

$$c(\lambda) = \int_{R^{-1}}^R \frac{ds}{2\pi i} s^{i\lambda - 1/2} \left(\tfrac{1}{2} \frac{1}{s - e^{i\alpha}} - \tfrac{1}{2} \frac{1}{s - e^{-i\alpha}}\right) + \mathcal{O}(R^{-1/2})\,. \qquad (5.15)$$

The integrand is meromorphic in the annular sector $\{s : 1/R < |s| < R,\ \arg(s) \in (0, 2\pi)\}$. To evaluate the integral, we consider the contour given in Fig. 2.
The integral over the circles which are concentric around the origin contributes $\mathcal{O}(R^{-1/2})$ and the integral over the segment $1/R \leq s \leq R$, $\arg(s) = 2\pi$ equals $(e^{2\pi i})^{i\lambda - 1/2} c(\lambda)$. Letting $R \to \infty$, we obtain

$$0 = -c(\lambda) + (e^{2\pi i})^{i\lambda - 1/2} c(\lambda)$$
$$+ \sum_{z \in \{e^{i\alpha}, e^{i(2\pi - \alpha)}\}} \operatorname*{Res}_{s=z} \left(\tfrac{1}{2} \frac{s^{i\lambda - 1/2}}{s - e^{i\alpha}} - \tfrac{1}{2} \frac{s^{i\lambda - 1/2}}{s - e^{-i\alpha}}\right).$$

This leads to

$$c(\lambda) = -\frac{\cosh\big((\pi - \alpha)\lambda - i\alpha/2\big)}{2\cosh(\pi\lambda)}\,.$$

Note that $c(\lambda) = 0$ when $\alpha = \pi$.

We next observe that on the space $L^2(\mathbf{R}^+) \oplus L^2(\mathbf{R}^+)$, we have

$$\mathcal{M}Q\mathcal{M}^* = \begin{pmatrix} 0 & c(\lambda) \\ c(\lambda) & 0 \end{pmatrix}.$$

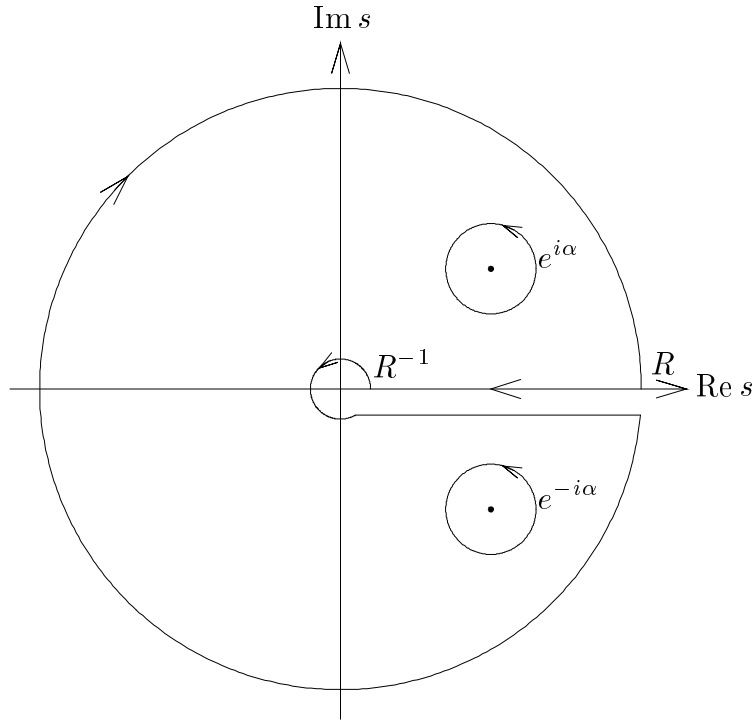

**Fig. 2**: The contour used in evaluating the integral $c(\lambda)$ of Eq.(5.15).

From this it follows at once that the spectrum of the corner contribution is

$$\sigma(Q) \;=\; \{\pm\frac{1}{2}\frac{\cosh(\lambda(\pi-\alpha)+i\alpha/2)}{\cosh(\lambda\pi)} \mid \lambda \in \mathbf{R}\}\;.$$

This set is shown in Fig. 3, for various values of $\alpha$ as a function of $\lambda$. It is easy to check that the set $\sigma(Q)$ is contained in the disk of radius $\frac{1}{2}\cos(\alpha/2)$.

The "diagonal part" $B^{\mathrm{sing}}$ is equal to $B^{\mathrm{sing}}(1-\sum_j \chi_j^{\mathrm{opposite}})$. In this case, both $s$ and $s'$ are on the same side of a corner, and we can reapply the bounds of Lemma 5.5. The proof of Proposition 5.4 is complete.

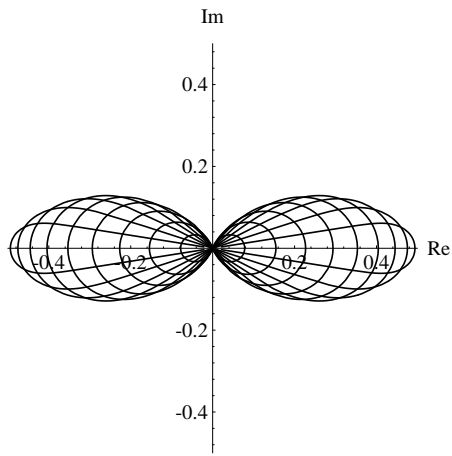

**Fig. 3**: The functions $c(\lambda)$ for angles $\alpha = \pi j/20$, $j = 1, \ldots, 19$.

## Appendix

The definition and explicit study of the operator $C_z$ are more complicated than those of $A_z$ given in [EP1] and those of $B_z$ given in Section 5. These additional difficulties are generated by the appearance of *two* derivatives in the definition of $C_z$, which make it an operator of order 1. However, when considering the integral kernel of $C_z$, important cancellations of singularities take place, *even in the presence of corners*, and the main purpose of this appendix is a sketch of the mechanisms implying these calculations. Detailed notes on this problem can be requested from the authors.

We fix first some simple notations. If $x, y \in \mathbf{R}^2$, then we denote

$$\begin{aligned}
r &= r(x,y) = |x-y|, \\
e &= e(x,y) = \nabla_x r(x,y) = -\nabla_y r(x,y) = (x-y)/r(x,y), \\
\dot{x}(s) &= T(s), \\
\dot{T}(s) &= -\varkappa(s)N(s) + \sum_j T_j \delta(s-s_j), \\
\dot{N}(s) &= \varkappa(s)T(s) + \sum_j N_j \delta(s-s_j).
\end{aligned}$$

The sums above are over the corners $j$, which are located at $s_j \in \Gamma$, where

$$T_j = \lim_{\delta \downarrow 0} T(s_j + \delta) - T(s_j - \delta),$$

and with a similar definition for $N_j$. The operator $C_z$ is defined starting from the double layer potential $\Phi_2 u$, which we defined for $x \notin \Gamma$ by

$$(\Phi_2 u)(x) = \int_\Gamma ds\, N(s) \cdot (\nabla_2 G_z)(x, x(s))u(s).$$

Then
$$(C_z u)(s) = \lim_{\varepsilon \downarrow 0} N(s) \cdot \nabla(\Phi_2 u)(x(s) + \varepsilon N(s)) ,$$

for all $s$ except the corners $s = s_j, j = 1, \ldots n$. It is not really obvious that the above limit exists, but we shall give a few arguments on the way to proving this. We want to argue only modulo Hilbert-Schmidt operators and we use the $\approx$ sign to denote equality modulo Hilbert-Schmidt operators (with bounds independent of the energy). We note first the identity, with $e = e(x, x(s'))$, $r = r(x, x(s'))$,

$$\nabla(\Phi_2 u)(x) = z \int_\Gamma ds' u(s') \Big\{ e(e \cdot N(s')) G(z^{1/2} r)$$
$$+ \Big( 2e(e \cdot N(s')) - N(s') \Big) \frac{G(z^{1/2} r)}{z^{1/2} r} \Big\} .$$

Here, again, $G(x) = \frac{i}{4} H_0^{(1)}(x)$. In the limit process, the arguments of $e$ and $r$ will be $(x(s) + \varepsilon N(s), x(s'))$, and then we find that (modulo Hilbert-Schmidt, as announced)

$$(C_z u)(s) \approx \lim_{\varepsilon \downarrow 0} \int_\Gamma ds' u(s') N(s) \cdot M_\varepsilon(s, s') N(s') z \frac{G(z^{1/2} r)}{z^{1/2} r}$$
$$\approx -\frac{1}{2\pi} \lim_{\varepsilon \downarrow 0} \int_\Gamma ds' u(s') N(s) \cdot M_\varepsilon(s, s') N(s') \frac{1}{r^2} ,$$

where $M_\varepsilon$ is the $2 \times 2$ matrix
$$2 |e\rangle\langle e| - 1 .$$

Observing that the trace of $M_\varepsilon$ is 0, one can operate a certain number of cancellations. Furthermore, it should be noted that the vectors $T_j$ and $N_j$ describing the change of angle at a corner are orthogonal. This implies a cancellation of the first *two* orders of the most singular contributions at the corners, through the identity

$$\lim_{\delta \to 0} N(s_j - \delta) \cdot M_\varepsilon \cdot T(s_j + \delta) = \lim_{\delta \to 0} T(s_j - \delta) \cdot M_\varepsilon \cdot N(s_j + \delta) .$$

(All these statements hold modulo Hilbert-Schmidt contributions.) A lengthy calculation, involving the formula

$$\partial_s \frac{e \cdot T(s')}{r} = -(1 + \varepsilon \varkappa(s)) \frac{T(s) \cdot M_\varepsilon(s, s') \cdot T(s')}{r^2}$$

shows, after integration by parts, that the result of these cancellations is

$$(C_z u)(s) \approx -\frac{1}{2\pi} \lim_{\varepsilon \downarrow 0} \frac{1}{1 + \varepsilon \varkappa(s)} \partial_s \int_\Gamma ds' u(s') \frac{e \cdot T(s')}{r}$$
$$\approx -\lim_{\varepsilon \downarrow 0} \partial_s \int_\Gamma ds' u'(s') G\Big( z^{1/2} r \big(x(s) + \varepsilon N(s), x(s')\big) \Big) ,$$

where $u'$ is the derivative of $u$.

This result also allows us to compare $C_z$ to $A_z$. In particular, it means that

$$C_z \approx i^{-1}\partial_s A_z i^{-1}\partial_s ,$$

and therefore, by the Structure Theorem II of [EP2], we see that for all $\alpha \in [0,1]$,

$$(1-\partial_s^2)^{-\alpha/2} C_z (1-\partial_s^2)^{\alpha/2-1/2} = \tfrac{1}{2} + \mathcal{B} + \mathcal{H} + \mathcal{T}_z ,$$

where $\|\mathcal{B}\| < \tfrac{1}{2}$, $\mathcal{H}$ is Hilbert-Schmidt and $\mathcal{T}_z$ is trace class. Note that $\mathcal{B}$ does not depend on $z$ and that $T_z$ is actually in the Schatten class $2/3$. Of course, these facts also follow from the identity Eq.(2.7) and from the (explicitly proved) facts about $A_z$, [EP2, Structure Theorem II] and $B_z$, Structure Theorem 5.1.

# References

[AS] Abramowitz, M. and I. Stegun: *Handbook of Mathematical Functions*, New York, Dover (1965).
[BS] Birman, M.Sh. and M.Z. Solomyak: *Spectral Theory of Selfadjoint Operators in Hilbert Space*, Dordrecht, Reidel (1987).
[BS2] Burmeister, B. and F. Steiner: (To appear).
[CH] Courant, R. and D. Hilbert: *Methods of Mathematical Physics*, Vol. 1, New York, J. Wiley (1953).
[DS] Doron, E. and U. Smilansky: Semiclassical quantization of billiards—a scattering approach. Nonlinearity 1055–1084 (1992).
[EP1] Eckmann, J.-P. and C.A. Pillet: Spectral duality for planar billiards. Commun. Math. Phys. **170**, 283–313 (1995).
[EP2] Eckmann, J.-P. and C.A. Pillet: Scattering phases and density of states for exterior domains. Ann. Inst. Henri Poincaré **62**, 383–399 (1995).
[Ha] Hairer, E.: (Private communication).
[He] Hesse, T.: Trace formula for a non-convex billiard system. Preprint.
[HS] Harayama, T. and A. Shudo: Zeta function derived from the boundary element method. Physics Lett. **A165**, 417–426 (1992).
[H] Hörmander, L.: *The Analysis of Linear Partial Differential Equations*, Berlin, Heidelberg, New York, Springer (1983–1985).
[JK] Jensen, A. and T. Kato: Asymptotic behaviour of the scattering phase for exterior domains. Comm. Part. Diff. Equ. **3**, 1165–1195 (1978).
[KR] Kleinman, R.E. and G.F. Roach: Boundary integral equations for the three-dimensional Helmholtz equation. SIAM Review **16**, 214–236 (1974).
[N] Nečas, J.: *Les Méthodes Directes en Théorie des Equations Elliptiques*, Paris: Masson (1967).
[R1] Robert, D.: Sur la formule de Weyl pour des ouverts non bornés. C. R. Acad. Sci. Paris **319**, 29–34 (1994).
[R2] Robert, D.: A trace formula for obstacles problems and applications. In *Mathematical Results in Quantum Mechanics. International Conference in Blossin (Germany), May 17–21, 1993. Series: Operator Theory: Advances and Applications* **70**, (Demuth, M., Exner, P., Neidhardt, H., Zagrebnov, V. eds.). Basel, Boston, Berlin, Birkhäuser (1994).
[SU] Smilansky, U. and I. Ussishkin: The smooth spectral counting function and the total phase shift for quantum billiards. J. Physics A (in print).
[VS] Vasil'ev, D.G. and Yu.G. Safarov: The asymptotic distribution of eigenvalues of differential operators. Amer. Math. Soc. Transl. (2) **150**, 55–110 (1992).